\documentclass[letterpaper, 10 pt, conference]{ieeeconf}

\usepackage[british]{babel}
\usepackage{amsmath,amssymb,amsfonts, bm}
\usepackage{graphicx}
\usepackage{textcomp}
\usepackage{float}
\usepackage{comment}

\usepackage{mathrsfs}
\usepackage[dvipsnames]{xcolor}
\usepackage[acronym]{glossaries}
\usepackage{lipsum}
\usepackage{pgfplots}
\usepackage{algorithm}
\usepackage{algpseudocode}

\newif\ifshowremove
\showremovetrue
\newcommand{\dec}{\mathrm{dec}}
\newcommand{\proj}{\mathrm{Proj}}% Remarks Leftherios

\newcommand{\Borel}{\mathcal{B}}

\newcommand{\X}{\mathbb{X}}

\newcommand{\true}{\top}

\newtheorem{theorem}{Theorem}

\newtheorem{definition}[theorem]{\noindent Definition}
\newtheorem{remark}{Remark}
\newtheorem{lemma}[theorem]{Lemma}
\newtheorem{assumption}{Assumption}
\usetikzlibrary{arrows.meta}

\def\mycmd{0} % 0 if arXiv version, 1 if ADHS version

\allowdisplaybreaks
\IEEEoverridecommandlockouts                           

\begin{document}

    \title{Risk-Aware Real-Time Task Allocation for\\Stochastic Multi-Agent Systems under STL Specifications \thanks{\hspace{-1.1em}This work is supported by the Dutch NWO Veni project CODEC, grant number 18244 and the European project SymAware, grant number 101070802. $^1$Department of Electrical Engineering (Control Systems Group), Eindhoven University of Technology, The Netherlands. Emails:\{m.h.w.engelaar, z.zhang3, m.lazar, s.haesaert\}@tue.nl. $^2$Division of Decision and Control Systems, School of Electrical Engineering and Computer Science, KTH Royal Institute of Technology, Sweden. Email: \{vlahakis, dimos\}@kth.se}}
    
    \author{M. H. W. Engelaar$^1$, Z. Zhang$^1$, E. E. Vlahakis$^2$, D.V. Dimarogonas$^2$, M. Lazar$^1$, and S. Haesaert$^1$}
    
    \maketitle

    \begin{abstract}
        This paper addresses the control synthesis of heterogeneous stochastic linear multi-agent systems with real-time allocation of signal temporal logic (STL) specifications. Based on previous work, we decompose specifications into sub-specifications on the individual agent level. To leverage the efficiency of task allocation, a heuristic filter evaluates potential task allocation based on STL robustness, and subsequently, an auctioning algorithm determines the definitive allocation of specifications. Finally, a control strategy is synthesized for each agent-specification pair using tube-based model predictive control (MPC), ensuring provable probabilistic satisfaction. We demonstrate the efficacy of the proposed methods using a multi-shuttle scenario that highlights a promising extension to automated driving applications like vehicle routing.
	\end{abstract}

    %%%%%%%%%%%%%%%%%%%%%%%%%%%%%%%%%%%%%%%%%%%%%%%

	\section{Introduction}

    Multi-agent systems are frequently employed in complex tasks that require close collaborations. Formal specifications, such as signal temporal logic (STL), are increasingly utilized to define complex collaborative task objectives~\cite{Rama2014}. The collaborative control of multi-agent systems satisfying STL specifications has been studied using mixed-integer programming~\cite{sun2022multi}, funnel functions~\cite{Liu2022}, and control barrier functions~\cite{lindemann2020barrier}. Nevertheless, a critical challenge for the practical implementation of these methods is their reliability in uncertain and dynamic environments. For systems subject to stochastic disturbances, it is important to evaluate the impact of these disturbances on the task accomplishment of individual agents so that the associated risks are restricted~\cite{Yang2023}.

\begin{figure}[htbp]
    \noindent
    \hspace*{\fill} 
    \begin{tikzpicture}[scale=1]
    
    \definecolor{sorange}{RGB}{255, 205, 153}
    \definecolor{sblue}{RGB}{0, 64, 128}
    
    \node[] (wh) at (0cm, 0cm) {\includegraphics[width=3.2cm]{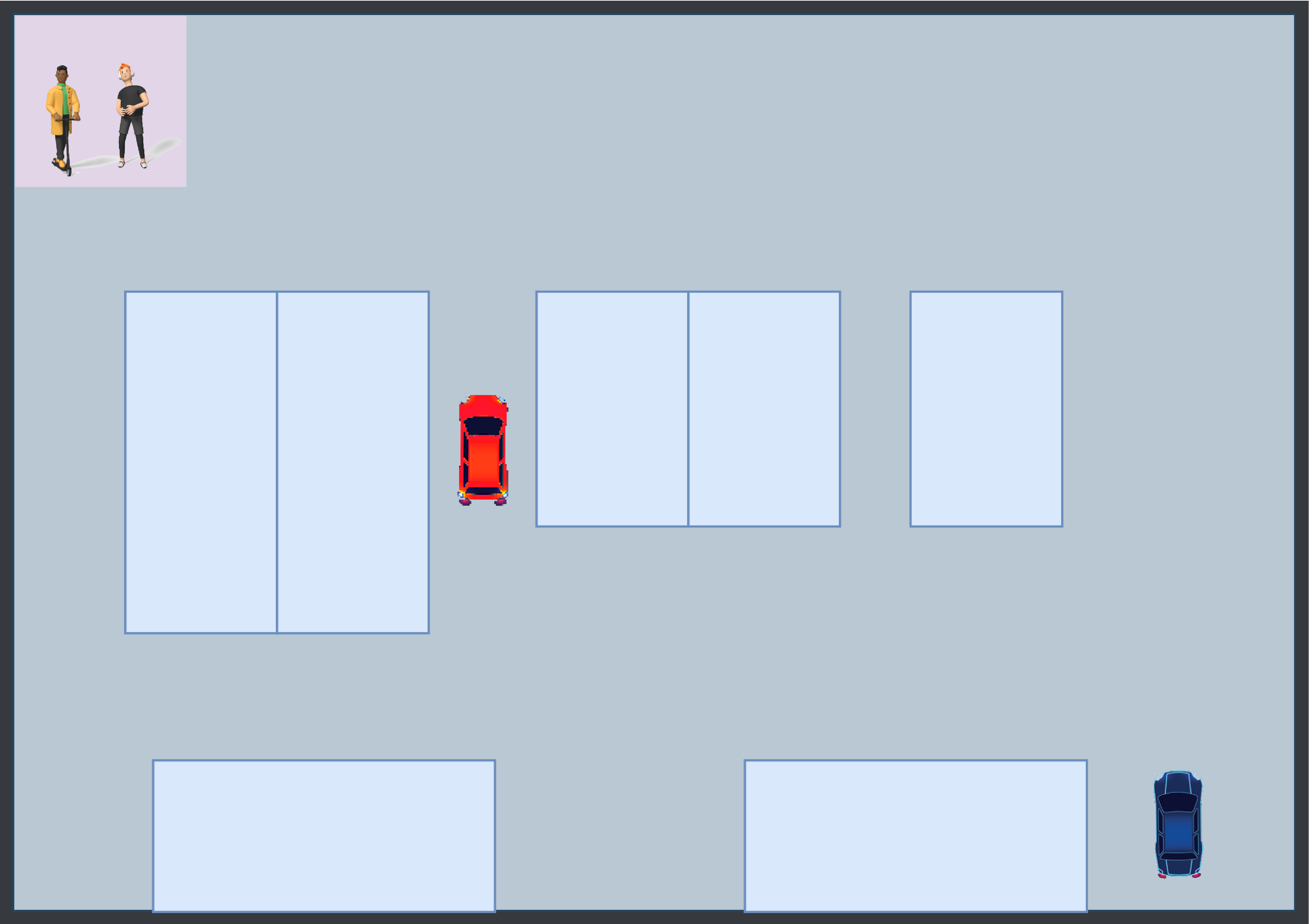}};
    
    \draw [-{Latex[length=2mm]}, very thick, sorange] plot [smooth, tension=1] coordinates { (-1.4cm, 0.7cm) (-1cm,-0.4cm) (0.2cm,-0.9cm)};
    
    \draw [-{Latex[length=2mm]}, very thick, sorange] plot [smooth, tension=1] coordinates { (-1.2cm, 0.9cm) (0cm,0.7cm) (1cm,0cm)};

    \node[] (wh1) at (4.5cm, 1cm) {\includegraphics[width=2.4cm]{warehouse.eps}};

    \draw [-{Circle[scale=0.5]}, thick, dashed, sblue] plot [tension=0] coordinates {(5.45cm, 0.5cm) (5.45cm,1.7cm) (3.6cm,1.7cm)};
    \draw [{Circle[scale=0.5]}-{Latex[length=1.2mm]}, thick, dashed, sblue] plot [tension=0] coordinates {(3.4cm, 1.5cm) (3.4cm,0.55cm) (4.3cm,0.55cm) (4.3cm,0.3cm) (4.67cm,0.3cm)};
    \draw [-{Circle[scale=0.5]}, thick, dashed, red] plot [tension=0] coordinates {(4.17cm, 2cm-0.85cm) (4.17cm,2cm-0.5cm) (3.5cm,2cm-0.5cm)};
    \draw [{Circle[scale=0.5]}-{Latex[length=1.2mm]}, thick, dashed, red] plot [tension=0] coordinates {(3.6cm,2cm-0.4cm) (5.35cm,2cm-0.4cm) (5.35cm,2cm-1cm) (5.25cm,2cm-1cm)};

    \node[] (wh2) at (4.5cm, -1cm) {\includegraphics[width=2.4cm]{warehouse.eps}};
    
    \draw [-{Circle[scale=0.5]}, thick, dashed, sblue] plot [tension=0] coordinates {(5.45cm, 0.5cm-2cm) (5.45cm,1.7cm-2cm) (3.6cm,1.7cm-2cm)};
    \draw [{Circle[scale=0.5]}-{Latex[length=1.2mm]}, thick, dashed, red] plot [tension=0] coordinates {(3.4cm, 1.5cm-2cm) (3.4cm,0.55cm-2cm) (4.3cm,0.55cm-2cm) (4.3cm,0.3cm-2cm) (4.67cm,0.3cm-2cm)};
    \draw [-{Circle[scale=0.5]}, thick, dashed, red] plot [tension=0] coordinates {(4.17cm, -0.85cm) (4.17cm,-0.5cm) (3.5cm,-0.5cm)};
    \draw [{Circle[scale=0.5]}-{Latex[length=1.2mm]}, thick, dashed, sblue] plot [tension=0] coordinates {(3.6cm,-0.4cm) (5.35cm,-0.4cm) (5.35cm,-1cm) (5.25cm,-1cm)};
    
    %%%%%
    \draw [-{Latex[length=4mm]}, line width=1mm] (wh.east) -- ([xshift=1.4cm] wh.east);
    
    \node[] at (2.4cm,0.25cm) {\scriptsize Decompose \&};
    \node[] at (2.3cm,-0.25cm) {\scriptsize Allocate};
    
    \node[] at (4.5cm,0cm) {\scriptsize OR};
    
    \end{tikzpicture}
    \hspace{\fill} 
    \caption{Decomposing and allocating tasks for vehicle routing.}
    \label{fig:structure}
    \end{figure}

    Despite recent efforts, a notable gap exists in the literature regarding the control of stochastic multi-agent systems under real-time allocated STL specifications with probabilistic guarantees. Such systems have significant practical value, exemplified by scenarios such as vehicle routing \cite{braekers2016vehicle}. For example, consider a scenario where two passengers require taxi rides, as illustrated in Fig.~\ref{fig:structure}. This naturally renders a task allocation problem by assigning each passenger a taxi. Finding the most efficient assignment is a critical challenge.

    %Dynamically assigned tasks have been described using real-time STL specifications~\cite{zhu2006neural, bruno2018dynamic}. NOTE: Neither references refers to STL!!

    Inspired by the example, this paper addresses 1) the problem of task allocation (TA) of STL specifications, which aims to determine an efficient agent-specification mapping in a multi-agent setting and 2) the control synthesis of agent-specification pairs assigned at runtime with probabilistic guarantees. TA necessitates decomposing specifications assigned to multiple collaborative agents into individual agent specifications, requiring the decomposition of STL specifications both among agents~\cite{charitidou2021signal, leahy2022fast} and across time~\cite{zhang2023modularized}. Previous work yields a linear programming problem~\cite{liu2015novel}, efficiently solvable using auction algorithms~\cite{bertsekas1992auction}. With decomposed real-time specifications, control synthesis for stochastic multi-agent systems holds promise with the utilization of shrinking-horizon model predictive control (MPC)~\cite{FarahaniTAC2019} to ensure bounded risks~\cite{mustafa2023probabilistic}. In previous work, probabilistic reachable tubes have been utilized to optimize risk bounds for both predefined specifications~\cite{Enge2023} and real-time specifications~\cite{Enge2024}. These methods underpin the risk-aware control synthesis of stochastic multi-agent systems with decomposed real-time specifications.

    This paper proposes a novel approach to allocating STL specifications, involving subgroups of heterogeneous agents and synthesizing control strategies for individual agents in real-time. By decomposing newly assigned specifications into agent-level tasks, we employ a heuristic approach based on STL robustness \cite{Donz2010} to assess agent-specification pairs and attack task assignment via an auctioning scheme formulated as a mixed integer linear program. Subsequently, a tube-based model predictive control strategy is employed where an optimization is solved at every time step at agent level. The solution of this optimization ensures the satisfaction of previous and newly allocated tasks with a desired risk level, while discarding tasks not adhering to the risk specification. We prove recursive feasibility of this MPC scheme, leading to an open-loop system with provable probabilistic guarantees and a closed-loop system for which probabilistic guarantees follow naturally. As trust is crucial in dynamic multi-agent systems, where agents and environment may often change \cite{burnett2011trust}, our contribution is proposing a novel concept of a trustworthy (correct-by-design) control algorithm for multi-agent systems with real-time task assignment.

    % Our contributions 

    % Contribution 1 (Task allocation): We design a task allocation method based on an auctioning algorithm for the decomposed specifications. Additionally, we designed a heuristic filtering method utilizing STL robustness on agent-specification pairs to improve the efficiency of task auctioning.
    % Contribution 2 (Control Synthesis): Decomposed specifications allow us to extend control synthesis to multi-agent systems. Additionally, it allows for recursive feasibility, Open-loop (proven), and closed-loop (hinted) probability satisfaction on the multi-agent level.
    % Concluding sentence: Overall, our paper contributes to the real-time allocation of STL specifications and the control synthesis of multi-agent systems while ensuring risk awareness.
    
	%%%%%%%%%%%%%%%%%%%%%%%%%%%%%%%%%%%%%%%%%%%%%%%
    
	\section{Preliminaries and Problem Statement} \label{Sec:ProSet}

	For a given probability measure $\mathbb P$ defined over a Borel measurable space $(\X,\mathcal{B}(\X))$, we denote the probability of an event $\mathcal{A} \in \Borel(\X)$ as $\mathbb{P}(\mathcal{A})$. In this paper, we will work with Euclidean spaces and Borel measurability. Further details on measurability are omitted, and we refer the interested reader to \cite{Bert1996}. Additionally, any half-space $H \subset \mathbb{R}^n$ is defined as $H:=\{x \in \mathbb{R}^n \mid g^Tx\geq b\}$ with $g \in \mathbb{R}^n$ and $b \in \mathbb{R}$. A polyhedron is the intersection of finitely many half-spaces, also denoted as $H:=\{x \in \mathbb{R}^n \mid Gx\geq b\}$ with $G \in \mathbb{R}^{q\times n}$ and $b \in \mathbb{R}^q$. A polytope is a bounded polyhedron. The 2-norm is defined by $||x||=\sqrt{x^Tx}$ with $x \in \mathbb{R}^n$. The identity matrix is denoted by $I_n \in \mathbb{R}^{n \times n}$. The vector of all elements one and zero are denoted by $\boldsymbol{1}_{n} \in \mathbb{R}^{n}$ and $\boldsymbol 0_n \in \mathbb{R}^{n}$, re-spectively. If the context is clear, they will be denoted by $\boldsymbol 1$ and $\boldsymbol 0$. For any finite set $I$, the cardinality is denoted by $|I|$. $\proj_X(Y)$ denotes the projection of elements in $Y$ onto $X$.

	%%%%%%%%%%%%%%

	\subsection{Multi-agent Systems}

	In this paper, we consider a finite ordered set $\mathcal{I}$ of heterogeneous agents where each agent $i \in \mathcal{I}$ has linear time-invariant dynamics with additive noise, given by
	\begin{equation} \label{Sys}
		x^i(k+1)=A^ix^i(k)+B^iu^i(k)+w^i(k),
	\end{equation}
	where $(A^i,B^i)$ is stabilizable. We impose that all agents have the same state and input dimensions. For the $i$-th agent, $x^i \in \mathbb{X}^i \subseteq \mathbb{R}^n$ is the state, $x^i(0) \in \mathbb{X}^i$ is an initial state, $u^i \in \mathbb{U}^i \subseteq \mathbb{R}^m$ is the input and $w^i \in \mathbb{R}^n$ is an independent, identically distributed (i.i.d.) noise disturbance with distribution $\mathcal{Q}^i_w$, i.e.,  $w^i(k) \sim \mathcal{Q}^i_w$, which can have infinite support. We will assume that the distribution has at least a known mean and variance, with the latter required to be strictly positive definite. Additionally, we impose that the distribution is central convex unimodal\footnote{$\mathcal{Q}^i_w$ is in the closed convex hull of all uniform distributions on symmetric compact convex bodies in $\mathbb{R}^n$ (c.f. \cite[Def. 3.1]{Dhar1976}).}. Finally, we define that any two disturbances $w^i(k)$ and $w^j(l)$ are independent.

	For each agent $i \in \mathcal{I}$, we require there is a local controller. We define the local controllers as a sequence of policies $$\boldsymbol f^i:= \{f^i_0, f^i_1,\dots\},$$ such that $f^i_k \!\!:\!\! \mathbb{H}^i_k\!\! \to \!\! \mathbb{U}^i$ maps the history of states and inputs to inputs. Here $\mathbb{H}^i_k\!:=\!(\mathbb{X}^i \times \mathbb{U}^i)^k \times \mathbb{X}^i$ with elements $\eta^i(k)\!:=\!(x^i(0),u^i(0), \cdots, u^i(k-1), x^i(k))$. By implement-ing controller $\boldsymbol f^i$ upon the corresponding agent, we obtain the controlled form of \eqref{Sys} for which the control input satisfies the feedback law $u^i(k)= f^i_k(\eta^i(k))$ with $\eta^i(k) \in \mathbb{H}^i_k$. We indicate the input sequence of agent $i$ by $\boldsymbol u^i:=\{u^i(0),u^i(1),\dots\}$ and define its executions as sequences of states $\boldsymbol x^i:=\{x^i(0),x^i(1),\dots\}$ referred to as signals. 
 
    We define the suffix, segment and signal element of any signal $\boldsymbol x^i:=\{x^i(0),x^i(1), \ldots\}$, respectively, by $\boldsymbol{x}^i_k=\{x^i(k), x^i(k+1), \dots\}$, $\boldsymbol x^i_{[a,b]}:=\{x^i(a), \ldots, x^i(b)\}$ with $a\leq b$, and $\boldsymbol x^i(k):=\boldsymbol x^i_{[k,k]}$. We define the concatenation of $\boldsymbol x^i$, for all agents $i \in I \subseteq\mathcal{I}$, by a higher dimensional signal $\boldsymbol x^I$, where $\boldsymbol x^I(k):=[\boldsymbol x^{i_1}(k)^T, \cdots, \boldsymbol x^{i_l}(k)^T]^T$, $i_j \in I$, and $l=|I|$ is the number of elements in $I$. Any signal $\boldsymbol{x}^i$ can be interpreted as a realization of the probability distribution induced by implementing controller $\boldsymbol{f}^i$, denoted by $\boldsymbol{x}^i \sim \mathbb{P}_{\boldsymbol{f}^i}$. Similar realizations exist for segments, i.e., $\boldsymbol{x}^i_k \sim \mathbb{P}_{\boldsymbol{f}^i,k}$  and concatenations, i.e., $\boldsymbol x^I \sim \mathbb{P}_{\boldsymbol f^I}$, with $\boldsymbol f^I:=\{\boldsymbol f^{i} \mid i \in I\}$.

	%%%%%%%%%%%%%%

	\subsection{Signal Temporal Logic \& Probabilistic Satisfaction}

	To mathematically describe tasks, e.g., reaching a target within a given time frame, we consider specifications given by signal temporal logic (STL). Here, we assume that all STL specifications adhere to the \emph{negation normal form} (NNF), see \cite{Rama2014}. This assumption does not restrict the overall framework since \cite{Sadr2015} shows that every STL specification can be rewritten into the negation normal form.

	STL consists of predicates $\mu^l$ for subgroups of $l$ agents that are either true ($\top$) or false ($\bot$). Each predicate $\mu^l$ is described by a function $h:\mathbb{R}^{nl} \to \mathbb{R}^q$, $l \leq |\mathcal{I}|$, as follows
	\begin{equation}
		\mu^l:=\begin{cases}
			\top \text{ if } \forall i \in \{1, \cdots, q\}, \ h_i(x) \geq 0 \\
			\bot \text{ if } \exists i \in \{1, \cdots, q\}, \ h_i(x) < 0.
		\end{cases}
	\end{equation}
    Here $h_i: \mathbb{R}^{nl} \to \mathbb{R}$ indicates the $i$-th element of function $h$. We will assume that all predicate functions are affine functions of the form $h(x)=Gx - b$ with $G \in \mathbb{R}^{q \times nl}$, $b\in \mathbb{R}^q$ and $||g_j||=1$ for all $j \in \{1, \cdots, q\}$, where $g_j$ is the $j$-th row of $G$. Furthermore, we require that $H_{\mu^l}:=\{ x \in \mathbb{R}^{nl} \mid G x \geq b\}$ describes a polytope. Let the set of all predicates be given by $\mathcal{P}$. The STL syntax will be given by
	$$\phi::= \true \mid \mu^l \mid \lnot \mu^l \mid \phi_1 \wedge \phi_2 \mid \phi_1 \vee \phi_2 \mid \square_{[a,b]} \phi \mid \phi_1 \ U_{[a,b]} \ \phi_2, $$
	where $\mu^l \in \mathcal{P}$, $\phi$, $\phi_1$ and $\phi_2$ are STL formula, $a,b \in \mathbb{N}$, and $a \leq b$. The semantics are given next, where $\boldsymbol{x}^{\mathcal{I}}_k \vDash \phi$ denotes the satisfaction of $\phi$ verified over the suffix of signal $\bm{x}^{\mathcal{I}}$.
	\begin{definition}
		The STL semantics, for a specification jointly assigned to a set of agents, are recursively given by:
		\begin{align*}
			&	 \boldsymbol{x}^{\mathcal{I}}_k \vDash \mu^l & \iff & \exists I \subseteq \mathcal{I}: \ |I|=l, \ h({\boldsymbol x^I(k)})\geq \boldsymbol 0  \\
			&	 \boldsymbol{x}^{\mathcal{I}}_k \vDash \lnot \mu^l & \iff & \nexists I \subseteq \mathcal{I}: \ |I|=l, \ h({\boldsymbol x^I(k)})\geq \boldsymbol 0  \\
			&	 \boldsymbol{x}^{\mathcal{I}}_k \vDash \phi_1 \wedge \phi_2 & \iff & \boldsymbol{x}^{\mathcal{I}}_k \vDash \phi_1 \text{ and } \boldsymbol{x}^{\mathcal{I}}_k \vDash \phi_2 \\
			&	 \boldsymbol{x}^{\mathcal{I}}_k \vDash \phi_1 \vee \phi_2 & \iff & \boldsymbol{x}^{\mathcal{I}}_k \vDash \phi_1 \text{ or } \boldsymbol{x}^{\mathcal{I}}_k \vDash \phi_2 \\
			&	 \boldsymbol{x}^{\mathcal{I}}_k \vDash \square_{[a,b]} \ \phi & \iff & \forall k' \in [k+a,k+b]: \boldsymbol{x}^{\mathcal{I}}_{k'} \vDash \phi \\
			&	 \boldsymbol{x}^{\mathcal{I}}_k \vDash \phi_1 \ U_{[a,b]} \ \phi_2 & \iff & \exists k_1 \in [k+a,k+b]: \boldsymbol{x}^{\mathcal{I}}_{k_1} \vDash \phi_2 \\
			&	&  &\hspace{-.75cm}  \text{ and }  \forall k_2 \in [k,k_1], \ \boldsymbol{x}^{\mathcal{I}}_{k_2} \vDash \phi_1.
		\end{align*}
	\end{definition}
	Here, we introduced a slight abuse of the notation as $[a,b]$ is used to describe the set of all natural numbers within the real interval $[a,b]$. Note that the definition is invariant of the actual indices of the $l$ agents. Additional operators can be derived such as the eventually-operator $\lozenge_{[a,b]} \phi := \true U_{[a,b]} \phi$.

	Since all agents behave stochastically, each STL specification can only be satisfied probabilistically. Should a specification be assigned at time $k$, the probability can be determined based on the state measurement $x^{\mathcal{I}}(k)$, the controller $\boldsymbol f^{\mathcal{I}}$, and the system dynamics \eqref{Sys}. Accordingly, we consider the probability that suffix fragment $\boldsymbol x^{\mathcal{I}}_k$ satisfies specification $\phi$, given state $x^{\mathcal{I}}(k)$. We denote this by
	  \begin{equation}\label{Eq:Multi_Prop}
		\mathbb{P}_{\boldsymbol f}(\phi,k):=\mathbb{P}_{\boldsymbol f^{\mathcal{I}},k}(\boldsymbol x^{\mathcal{I}}_k \vDash \phi \mid x^\mathcal{I}(k)).
	\end{equation}
    For each specification $\phi$, we require that the maximal allowable risk $r_{\phi, \max}$ of not satisfying specification $\phi$ is given. In the remainder, we will refer to this as the maximal risk. Note that for $|\mathcal{I}|=1$, the above definitions of STL and probability will become the standard ones as seen in \cite{Enge2024}.

	%%%%%%%%%%%%%%

	\subsection{Problem Statement \& Approach} \label{Sec:ProbSta}

	Consider each agent $i \in \mathcal{I}$ has dynamics \eqref{Sys} with noise distribution $\mathcal{Q}^i_w$ and state update $x^i(k)$, at each time step $k \in \mathbb{N}$. We assume that a new specification may be provided at any time. The objective is to update the suffix controller $\boldsymbol f^{\mathcal{I}}_k$ at each time step $k \in \mathbb{N}$. Here, any newly provided specification is either accepted or rejected based on the maximal risks of the newly provided specification and previously accepted specifications. 
    \if\mycmd0
	An illustrative example can be found in the Appendix to further showcase the paper's objective. We consider the following problem statement. \smallskip
	\else
    An illustrative example can be found in the extended report \cite{Enge2024riskawaremulti} to further showcase the paper's objective. We consider the following problem statement. \smallskip
	\fi

	\noindent \textbf{Problem Statement.} 
    Given a sequence of STL specifications $\boldsymbol{\phi}:=\{\phi_1, \cdots, \phi_t\}$ assigned, respectively, at time instances $\{k_1, \cdots, k_t\}$ with $k_i<k_j$ for $i<j$, with maximal risk $r_{\phi_i,\max}$, develop a method for \emph{updating}, at each time step $k\in \mathbb{N}$, suffix $\boldsymbol f^{\mathcal{I}}_k$ such that $\mathbb{P}_{\boldsymbol f}(\phi_i, k_i)\geq1-r_{\phi_i,\max}$ if $k_i\leq k$ and $\phi_i$ is a previously accepted specification, where $\phi_i$ with $k_i=k$ is allowed to be rejected.
    
    \smallskip
    
	\noindent \textbf{Approach.}
	First, a newly provided STL specification and maximal risk are decomposed onto the single-agent level (Section \ref{Sec:Decomp}). Subsequently, a filter determines which agent-specification pairs are favorable based on STL robustness and heuristics (Section \ref{Sec:Filter}). Afterwards, for each remaining agent-specification pair, a potential control strategy is determined together with the local risk value (Section \ref{Sec:Control}). Based on the local risk value for each feasible agent-specification pair, auctioning determines the definitive assignment (Section \ref{Sec:Auction}). Should no valid assignment exist, the specification is rejected. Finally, the control strategy for each agent is updated and implemented (Section \ref{Sec:Implem}).

	\begin{figure}[htp]
        \vspace{-0.5em}
        \centering
		\includegraphics[width=0.8\columnwidth]{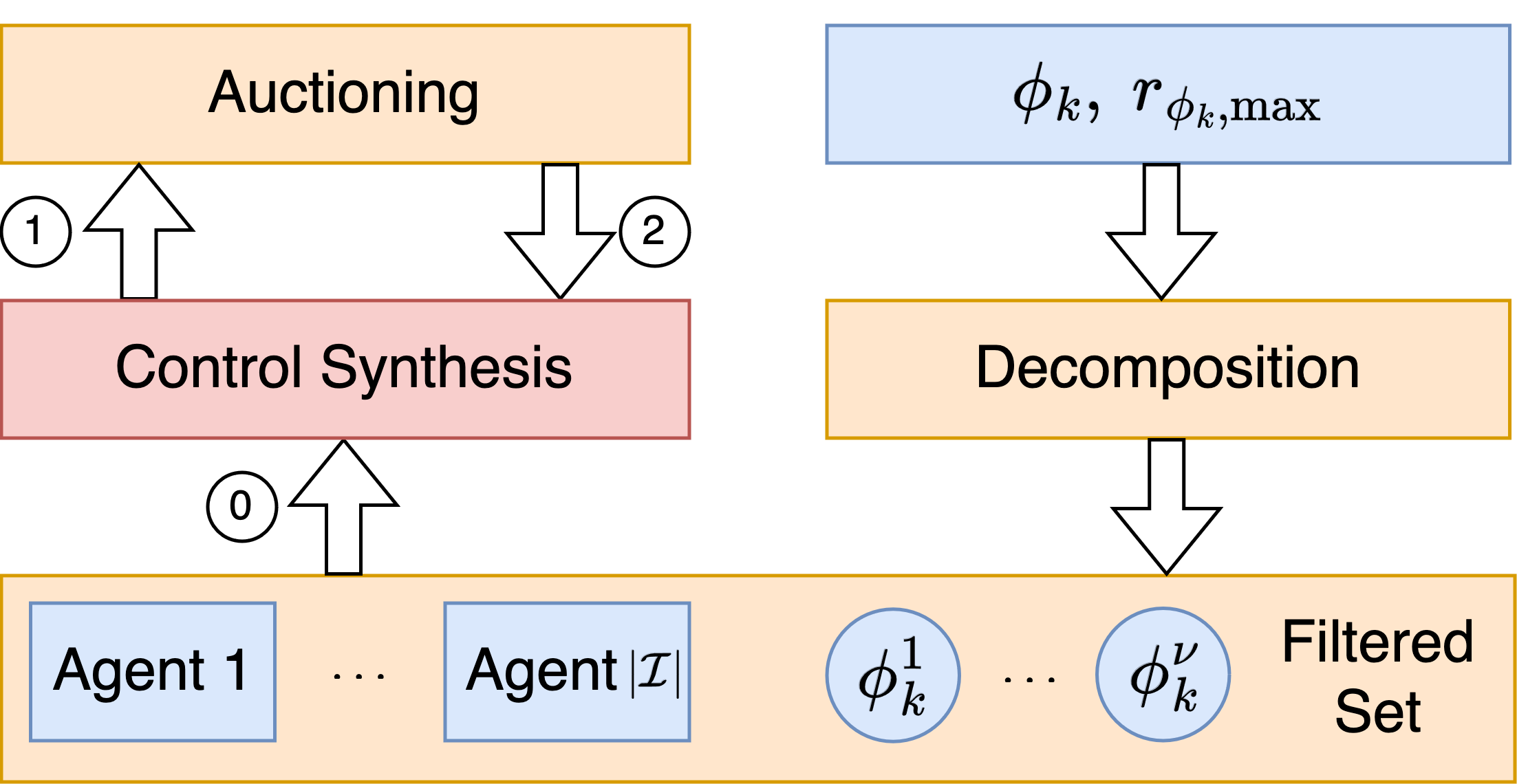}
        \vspace{-0.5em}
        \caption{Illustration of the approach at time $k$. (0) Determine for each agent-specification pair a control strategy and the local risk value. (1) Auctioning offers from agent-specification pairs. (2) Decision after auctioning. %Red is covered in Section \ref{Sec:SingleAgent} and orange in Section \ref{Sec:Main}. 
        }
		\label{Fig:Approach1}
        \vspace{-0.5em}
	\end{figure}

    %%%%%%%%%%%%%%%%%%%%%%%%%%%%%%%%%%%%%%%%%%%%%%%

	\section{Risk-Aware Multi-Agent Task Allocation and Control Synthesis} \label{Sec:Main}

    To tackle the problem above, we will first consider agent dynamic decomposition, specification decomposition and heuristic agent-specification filtering. Second, we establish control synthesis and local risk value computation for any individual agent-specification pair. Third, we utilize auctioning, based on the local risk values, to obtain a definitive agent-specification assignment. This section concludes with an implementation algorithm and theoretical analysis of the overall method.

    \subsection{Decomposing Agent Dynamics}

    We decompose the dynamics \eqref{Sys}, given by
    \begin{equation*} 
		x^i(k+1)=A^ix^i(k)+B^iu^i(k)+w^i(k),
	\end{equation*}
    into a nominal and an error part \cite{Hewi2018}. The nominal dynamics, denoted as $z^i$,  contain no stochasticity, and the (stochastic) error dynamics, denoted as $e^i$, are autonomous. This yields
	\begin{subequations} \label{Eq:NomErr}
		\begin{align}
			z^i(k+1) & = A^iz^i(k)+B^iv^i(k), \label{Eq:Nom}\\
			e^i(k+1) &= A^i_Ke^i(k)+w^i(k), \label{ErrorPart}\\
			\mbox{with  }
			x^i(k) &= z^i(k)+ e^i(k),\\
			u^i(k) &= v^i(k) +K^ie^i(k).
		\end{align}
	\end{subequations}
	Here $A^i_K=A^i+B^iK^i$ and $K^i$ is a stabilizing feedback gain meant to keep the error $e^i$ small. Throughout this section, we omit superscripts indicating agent indices when appropriate.

    %%%%%%%%%%%%%%

    \subsection{Specification Decomposition} \label{Sec:Decomp}
    
    Let $\phi$ be a newly assigned STL specification with maximal risk $r_{\phi,\max}$, involving a distinct number of agents, $\nu\leq |\mathcal{I}|$. The objective is to decompose the specification $\phi$ into agent-level sub-specifications, $\phi^j$, $j\in\mathcal{I}$, and allocate individual maximal risks, $r_{\phi^j,\max}$, in accordance to $r_{\phi,\max}$. First, we define $\dec(\phi,\nu)=\wedge_{j=1}^\nu\phi^j$ as the decomposition of $\phi$ into a conjunctive form, with each conjunct representing a sub-specification, $\phi^j$, the satisfaction of which depends on a single agent $j$, and $\nu$ being the number of agents needed to satisfy $\phi$. Second, we assign a maximal risk to $\phi^j$, following a union-bound argument, which requires $\sum_{j=1}^\nu r_{\phi^j,\max}\geq r_{\phi,\max}$. In this paper, we select a uniform allocation of risks across sub-specifications, obtaining $r_{\phi^j,\max}=\frac{r_{\phi,\max}}{\nu}$, for all agents $j\in \{1,\ldots,\nu\}$.
    
    The decomposition $\dec(\phi,\nu)$ acts as a transformation on $\phi$ returning a conjunctive formula, the satisfaction of which implies satisfaction of $\phi$. However, the converse is not necessarily true, implying a lack of completeness. There are a few methods \cite{charitidou2021signal,leahy2022fast,leahy2023rewrite} for splitting STL specifications into sub-specifications. In this paper, given a specification $\phi$, we convert Boolean and temporal operators to conjunctive forms based on the heuristics in \cite{leahy2023rewrite} and decompose predicates following a similar approach to \cite{charitidou2021signal}. Therein, predicates are defined on general convex sets, whereas negated forms are not studied. Here, we focus on polytopic settings and allow negations. %, for which we provide a decomposition procedure.     
    % This paper considers the rewrite-based approach in \cite{leahy2023rewrite}, which fits our multi-agent setting. Given a global STL specification, the decomposition therein results in a conjunctive formula containing sub-specifications, each assigned to a subteam of agents. 
    For our setting, we assume the following.
    \begin{assumption}\label{assum:decomposed_formula}
        Let $\phi$ be an STL specification assigned at time $k$. There exists
        decomposition $\dec(\phi,\nu)=\wedge_{j=1}^{\nu}\phi^j$, where $\phi^j$ indicates a task involving a single agent.
    \end{assumption}
    
    In view of existing decomposition approaches, where sub-specifications are assigned to subteams of agents, Assumption \ref{assum:decomposed_formula} is restrictive. However, in the setting of polytopic predicates, we can produce agent-level sub-specifications by projecting the associated predicates onto the agents' state space. Before explaining this, we define the following.
    \begin{definition}\label{def:orthotopes}
        Let $H:=\{x \mid h(x)\geq \boldsymbol 0\}$ be a polytope in $\mathbb{R}^n$. $B_{\mathrm{in}}(H):=\{x\mid \underline{b}_j^{\mathrm{in}}\leq x_j\leq \overline{b}_j^{\mathrm{in}},\; j=1,\ldots,n\}$ is an orthotope inscribed in $H$, i.e., $B_{\mathrm{in}}(H)\subseteq H$, and $B_{\mathrm{out}}(H):=\{x\mid \underline{b}_j^{\mathrm{out}}\leq x_j\leq \overline{b}_j^{\mathrm{out}},\; j=1,\ldots,n\}$ is an orthotope circumscribing $H$, i.e., $B_{\mathrm{out}}(H)\supseteq H$.
    \end{definition}
    
    The following lemma shows that any (negated) predicate can be decomposed into a conjunction of agent-level sub-predicates, of which satisfaction implies the satisfaction of the original (negated) predicate. The decomposition of the Boolean and temporal operators is described in \cite{leahy2023rewrite}. 
    
    \begin{lemma}\label{lemma:decomposition}
        Let $\mu:=(h(x)\geq \boldsymbol 0)$ be a predicate, $H:=\{x \mid h(x)\geq \boldsymbol 0\}$ be a polytope, with $x=[x_1^\intercal\;\ldots\;x_\nu^\intercal]^\intercal$, $x_i\in\mathbb{X}^{i}$, $i\in \{1,\ldots,\nu\}$, and $B_{\mathrm{in}}(H)\subseteq H$ and $B_{\mathrm{out}}(H)\supseteq H$ be the associated orthotopes inscribed in and circumscribing $H$, respectively. Then, it holds that (i) $\dec(\mu,\nu)=\wedge_{i=1}^\nu \mu_{\mathrm{in}}^i$ satisfies $\mu$, with $\mu_{\mathrm{in}}^i:=(x_i\in \proj_{\mathbb{X}^i}(B_{\mathrm{in}}(H))$, and (ii) $\dec(\neg\mu,\nu)=\wedge_{i=1}^\nu \neg\mu_{\mathrm{out}}^i$ satisfies $\neg \mu$, with $\mu_{\mathrm{out}}^i:=(x_i\in \proj_{\mathbb{X}^i}(B_{\mathrm{out}}(H))$.
    \end{lemma}
    
    \begin{proof}
        (i) Let $\mu_{\mathrm{in}}:=(x\in B_{\mathrm{in}}(H))$. By construction, it holds that $B_{\mathrm{in}}(H)\equiv \proj_{\mathbb{X}^1}(B_{\mathrm{in}}(H))\times \cdots\times \proj_{\mathbb{X}^\nu}(B_{\mathrm{in}}(H))$, implying that $\mu_{\mathrm{in}}\equiv \wedge_{i=1}^\nu\mu_{\mathrm{in}}^i$. Since, $B_{\mathrm{in}}(H)\subseteq H$, $\mu_{\mathrm{in}}$ satisfies $\mu$, which leads to $\wedge_{i=1}^\nu\mu_{\mathrm{in}}^i$ satisfying $\mu$ as needed. (ii) Let $\mu_{\mathrm{out}}:=(x\in B_{\mathrm{out}}(H))$. Similarly to (i), it holds that $\mu_{\mathrm{out}}\equiv \wedge_{i=1}^\nu\mu_{\mathrm{out}}^i$, since $B_{\mathrm{out}}(H)\equiv \proj_{\mathbb{X}^1}(B_{\mathrm{out}}(H))\times \cdots\times \proj_{\mathbb{X}^\nu}(B_{\mathrm{out}}(H))$, by construction. Then, we have $\neg \mu_{\mathrm{out}}\equiv\lor_{i=1}^{\nu}\neg\mu_{\mathrm{out}}^i$. Since $\wedge_{i=1}^{\nu}\neg\mu_{\mathrm{out}}^i$ satisfies $\lor_{i=1}^{\nu}\neg\mu_{\mathrm{out}}^i$ and $\neg\mu_{\mathrm{out}}$ satisfies $\neg \mu$, we have that $\wedge_{i=1}^{\nu}\neg\mu_{\mathrm{out}}^i$ satisfies $\neg \mu$, completing the proof. 
    \end{proof}
    \if\mycmd0
	An illustrative example of the specification decomposition method can be found in the Appendix.
	\else
    An illustrative example of the specification decomposition method can be found in the extended report \cite{Enge2024riskawaremulti}.
	\fi
    
    % Let $\mu:=(h(x)\geq 0)$ be a predicate, $H:=\{x \mid h(x)\geq 0\}$ be a polyhedron in $\mathbb{R}^n$, with $x=[x_1^\intercal\;\ldots\;x_\nu^\intercal]^\intercal$, $x_i\in\mathbb{R}^{n_i}$, $i=1,\ldots,\nu$, $\mathbb{B}_{\mathrm{in}}(H)$ be the maximum infinity norm ball inscribed in $H$, and $\mathbb{B}_{\mathrm{out}}(H)$ be the minimum infinity norm ball circumscribing $H$. Let $H^i_{\mathrm{in}}:=\{x_i \mid h_{\mathrm{in}}^i(x_i)\geq 0\}=\proj_{x_i}(\mathbb{B}_{\mathrm{in}}(H))$ be the projection of $\mathbb{B}_{\mathrm{in}}(H)$ onto the space of $x_i$, and $H^i_{\mathrm{out}}:=\{x_i \mid h_{\mathrm{out}}^i(x_i)\geq 0\}=\proj_{x_i}(\mathbb{B}_{\mathrm{out}}(H))$ be the projection of $\mathbb{B}_{\mathrm{out}}(H)$ onto the space of $x_i$. Then, $\dec(\mu,\nu)=\wedge_{i=1}^\nu \mu_{\mathrm{in}}^i$, where $\mu_{\mathrm{in}}^i:=(h_{\mathrm{in}}^i(x_i)\geq 0)$, and $\dec(\neg\mu,\nu)=\wedge_{i=1}^\nu \neg\mu_{\mathrm{out}}^i$, where $\mu_{\mathrm{out}}^i:=(h_{\mathrm{out}}^i(x_i)\geq 0)$, are valid decompositions.

    %%%%%%%%%%%%%%

    \subsection{Heuristic Agent-Specification Filtering} \label{Sec:Filter}

    STL is equipped with robustness metrics for assessing the robust satisfaction of a formula \cite{Donz2010}. A scalar-valued function $\rho^\phi(\bm{x},k)$ of a signal $\bm{x}$ and time $k$ indicates how robustly a signal $\bm{x}$ satisfies a formula $\phi$ at time $k$. The robustness function is defined recursively in \cite[Definition 3]{Donz2010}.
    
    % \begin{align*}
    %     \rho^\mu(\bm{x},k) &= \min(h(\boldsymbol x(k))), \ \rho^{\lnot \mu }(\bm{x},k) = -\rho^{\mu}(\bm{x},k),\\ 
    %     \rho^{\phi_1 \land \phi_2}(\bm{x},k) &= \min(\rho^{\phi_1}(\bm{x},k),\rho^{\phi_2}(\bm{x},k)), \\
    %     \rho^{\phi_1 \lor \phi_2}(\bm{x},k) &= \max(\rho^{\phi_1}(\bm{x},k),\rho^{\phi_2}(\bm{x},k)), \\
    %     \rho^{\square_{[a,b]}\phi}(\bm{x},k) &= \min_{k'\in [k+a,k+b]} \rho^{\phi}(\bm{x},k'), \\
    %     \rho^{\phi_1 U_{[a,b]}\phi_2}(\bm{x},k) &= \max_{k_1 \in [k+a,k+b]}\left( \min(R_1,R_2) \right),\\ \text{with }R_1&=\rho^{\phi_2}(\bm{x}, k_1), \ R_2=\min_{k_2\in [k,k_1] }\rho^{\phi_1}(\bm{x},k_2),
    % \end{align*}
    % with $\mu$ being a predicate, and $\phi$, $\phi_1$, and $\phi_2$ STL formulas.}

    Let $\phi_k$ be a newly assigned STL specification at time $k$, and assume that $\dec(\phi_k,\nu_k)=\wedge_{j=1}^{\nu_k}\phi_k^j$. In the following, we require that the total number of agents, denoted by $|\mathcal{I}|$, satisfies $|\mathcal{I}|>\nu_k$, otherwise, the following filtering procedure is ignored. Although, $\phi_k$ is split into $\nu_k$ agent-level sub-specifications, the assignment of $\phi_k^j$ to a specific agent $i\in\mathcal{I}$ is not determined by $\dec(\phi_k,\nu_k)$. In fact, each sub-specification, $\phi_k^j$, may be assigned to any agent $i\in\mathcal{I}$ resulting in $\nu_k|\mathcal{I}|$ agent-specification pairs. To reduce the number of potential pairs, we utilize the following heuristic approach: Based on the quantitative semantics of STL \cite{Donz2010}, we compute the robustness function of $\phi_k^j$ over the $i$th agent's trajectory. This metric indicates roughly the distance between the trajectory of agent $i$ and the space of signals that satisfy $\phi_k^j$ \cite{Fainekos2009},  thereby facilitating a preliminary agent-specification assessment. We approximate the $i$th agent's trajectory by $\bm{z}^{i,k-1}=\{z_0^{i,k-1},z_1^{i,k-1},\ldots\}$, $i\in \mathcal{I}$, that is, the signal associated with the (deterministic) nominal dynamics of agent $i$, \eqref{Eq:Nom}, computed at time $k-1$, which is available at time $k$. Specifically, for each $j\in\{1,\ldots,\nu_k\}$, and $i\in\mathcal{I}$, we efficiently calculate $\rho^{\phi_k^j}(\bm{z}^{i,k-1},k)$, which requires negligible computational cost. Subsequently, for each $\phi_k^j$, we collect the largest $\nu_k$ values of $\rho^{\phi_k^j}(\bm{z}^{i,k-1},k)$ among the $|\mathcal{I}|$ options, resulting in $(\nu_k)^2$ agent-specification pairs. Hence, should $v_k\ll|\mathcal{I}|$, the filtering significantly reduces potential pairs.
    
    %Finally, for $i\in\mathcal{I}$, we construct a tuple $(i,\Phi_k^i)$, where $\Phi_k^i$ collects all sub-specifications, $\phi_k^j$, that may be assigned to agent $i$. We denote by $\phi_k^j\in \Phi_k^i$ a sub-specification $\phi_k^j$ that may potentially be assigned to the $i$-th agent at time $k$. We write $\Phi_k^i=\emptyset$, when agent $i$ is not assigned any new sub-specification at time $k$.

    %%%%%%%%%%%%%%

    \subsection{Control Synthesis and Local Risk Value} \label{Sec:Control}

    In this subsection, we compute the local risk value and synthesize a controller for a given agent-specification pair utilizing previous work on risk-aware MPC for single-agent systems \cite{Enge2024}. We refer the reader to \cite{Enge2024} for more details.
    
    We assume a new specification $\phi$ is provided at time $k$ together with maximal risk $r_{\phi, \max}$. For any specifications $\psi$, we denote with $k_{\psi}$ the time at which the specification was assigned. We denote with $\mathcal{P}^k$ the set of all previously accepted specifications $\psi$ with $k_{\psi}< k$, together with the newly assigned specification $\phi$, i.e., $\phi \in \mathcal{P}^k$. The objective is to synthesize a controller $\boldsymbol{f}$ such that $\varphi:=\wedge_{\psi \in \mathcal{P}^k}\lozenge_{[k_{\psi}, k_{\psi}]} \psi$ is satisfied in accordance to the maximal risk of each conjunction element, i.e., $\mathbb{P}_{\boldsymbol{f}}(\psi,k_\psi)\geq 1-r_{\psi,\max}, \ \forall \psi \in \mathcal{P}^k$. To accomplish the objective, we make the following assumption.
    \begin{assumption}
        Distribution $\mathcal{Q}_w$ has zero mean and $A_K\Sigma_{\infty}A_K^T+\text{var}(\mathcal{Q}_w)=\Sigma_{\infty}$ has solution $\Sigma_{\infty}=I_n$
    \end{assumption}
    Note that the above assumption is not restrictive as a simple coordinate transformation $y=\Sigma_{\infty}^{-\frac{1}{2}}x$ will ensure the second part of the assumption is satisfied. We call the transformed dynamics the \textit{normalized dynamics}, and in the remainder, we maintain our notations as if the dynamics are normalized.
    
	We are interested in implementing a tube-based MPC algorithm that at each time instance $k\in[0,N]$ recomputes the optimal nominal trajectory $\boldsymbol{z}_{[k, N]}$, the optimal nominal input $\boldsymbol{v}_{[k,N-1]}$, and the optimal risk levels $\boldsymbol{r}_{[k,N]}$. For this, we assume given $x(k)$ together with $\boldsymbol z_{[0,k]}$, $\boldsymbol v_{[0,k-1]}$, $\boldsymbol \rho_{[0,k]}$ and $\boldsymbol r_{[0,k]}$ computed at time $k-1$. Furthermore, for all $\psi\in \mathcal{P}^k$, we let $\mathcal{H}_{\psi}^{k_{\psi}} \subseteq [0, N]$ without loss of generality. Here, the active horizon $\mathcal{H}_{\phi}^k$ is the set of all time instances needed to evaluate $\boldsymbol x_k \vDash \phi$ (see \cite[Definition 4]{Enge2024}). It should be noted that any active horizon consists of only finite elements.

	Let us consider the following tube-based MPC problem
	\begin{subequations}\label{Eq:TMPC}
		\begin{align}
			\min_{\Xi}  \   J(&\boldsymbol z_{[0,N]}, \boldsymbol v_{[0,N-1]}, \boldsymbol r_{[0,N]}), \label{Eq:TMPC_cost}\\
			\text{s.t. }	\ z(k) &= x(k) \label{Eq:TMPC_meas}\\
			z(\tau+1) &= Az(\tau)+Bv(\tau), \forall \tau \in \{k, \cdots, N-1\},\!\! \label{Eq:TMPC_sys} \\[0.4em]
			G(j)z(j)&+MF(j)s(k) \geq b(M,j) + \rho(j)\boldsymbol{1}, \label{Eq:TMPC_SpecOld}\\
			Cs(k) &\geq d, \ \epsilon \leq \rho(j)\leq \textstyle \frac{M}{2},\ \forall  j\in \{0, \cdots, N\}, \label{Eq:TMPC_SpecCon}\\[0.4em]
			n &= \rho(j)^2 r(j), \hspace{1.3cm}\forall  j\in \{0, \cdots, N\}, \label{Eq:TMPC_ConRiskPRT}\\
			r_{\psi,\max} &\geq \textstyle \sum_{j \in \mathcal{H}_\psi^{k_\psi}\setminus \{k_\psi\}} r(j),  \forall \psi \in  \mathcal{P}^k,  \label{Eq:TMPC_ProbGuar}
		\end{align}
	\end{subequations}
	where $\Xi:=\{\boldsymbol{z}_{[k, N]}, \boldsymbol{v}_{[k,N-1]}, \boldsymbol{r}_{[k+1,N]}, \boldsymbol \rho_{[k+1,N]}, s(k)\}$,  $J$  is either a linear or quadratic cost function, the constraints \eqref{Eq:TMPC_SpecOld}-\eqref{Eq:TMPC_SpecCon} are obtained from $\varphi=\wedge_{\psi \in \mathcal{P}^k}\lozenge_{[k_{\psi}, k_{\psi}]} \psi$ and the constraints \eqref{Eq:TMPC_ConRiskPRT}-\eqref{Eq:TMPC_ProbGuar} are obtained from \cite[Thm. 4]{Enge2024}. From the above tube-based MPC, an (updated) control strategy is obtained by $\boldsymbol v_{[k,N-1]}$, and a local risk value $r_{\phi,mpc}$ is obtained by the right-hand side of \eqref{Eq:TMPC_ProbGuar}, i.e.,
    \begin{equation}\label{Eq:RiskVal}
        \textstyle r_{\phi,mpc}=\sum_{j \in \mathcal{H}_\phi^{k}\setminus \{k\}} r(j).
    \end{equation}
    
    A solution to \eqref{Eq:TMPC}, by \cite[Thm. 4]{Enge2024}, ensures that for all $\psi \in \mathcal{P}^k$, $\mathbb{P}_{\boldsymbol{f}}(\psi,k_\psi)\geq 1-r_{\psi,\max}$. Should there exist no solution to problem \eqref{Eq:TMPC}, the newly provided agent-specification pair is deemed infeasible. In such a case, we will update the control strategy based only on measurement $x(k)$, as per constraint \eqref{Eq:TMPC_meas}. This can be achieved by considering the MPC problem \eqref{Eq:TMPC} with $\mathcal{P}^k=\mathcal{P}^k\setminus\{\phi\}$. However, should the measurement also fail to produce a controller, the previously computed control strategy will be implemented instead.
    
    %%%%%%%%%%%%%%
    
    \subsection{Task Allocation Algorithm} \label{Sec:Auction}

    Let $\phi$ be a specification which can be decomposed into $\nu$ sub-specifications. After filtering agent-specification pairs and attempting optimization problem \eqref{Eq:TMPC} with the remainder, a list $\mathbb{L} \subset \mathcal{I} \times \{1, \cdots, \nu\}$ of feasible agent-specification pairs is generated. Each feasible agent-specification pair has a local risk value \eqref{Eq:RiskVal} denoted by $r^i_{\phi_j,mpc}$, with $i \in \mathcal{I}$ and $j \in \{1, \cdots, \nu\}$. Our objective is to auction sub-specifications to agents based on the local risk values, obtaining a definitive assignment in the process. Here, each agent can only receive one specification, and all specifications must be allocated.
    
    To determine an agent-specification assignment, we solve the following binary optimization problem given by
    \begin{subequations}\label{Eq:OptiPair}
		\begin{align}
			\textstyle \min_{\boldsymbol \lambda}  \ \sum_{(i,j) \in \mathbb{L}} &\lambda_{i,j} r^i_{\phi_j,mpc},\\
			\textstyle \text{s.t. } \sum_{i \in \mathcal{I}} \lambda_{i,j} &= 1, \ \forall j \in \{1, \cdots, \nu\} \label{Eq:OptiPairAgent}\\
            \textstyle \sum_{j \in \{1, \cdots, \nu\}} \lambda_{i,j} &\leq 1, \ \forall i \in \mathcal{I}, \label{Eq:OptiPairSpec}\\
            \lambda_{i,j} &= 0, \ \ \ \ \ \  \forall (i,j) \notin \mathbb{L},\\
            \lambda_{i,j} &\in \{0,1\}, \ \forall (i,j) \in \mathbb{L},
		\end{align}
	\end{subequations}
    where $\boldsymbol \lambda =\{\lambda_{i,j} \mid (i,j) \in \mathbb{L}\}$. Here, constraint \eqref{Eq:OptiPairAgent} entails that each specification must be assigned to exactly one agent; constraint \eqref{Eq:OptiPairSpec} entails that each agent can at most receive one specification; and $\lambda_{i,j}=1$ implies agent $i$ and sub-specification $j$ are chosen as an agent-specification pair. The cost minimizes the sum of the local risk values, thereby ensuring the best probability satisfaction of specification $\phi$. Should no solution be found, specification $\phi$ will be rejected.

    %%%%%%%%%%%%%%

    \subsection{Implementation \& Theoretical Analysis} \label{Sec:Implem}

    To enact the proposed method, we consider Algorithm \ref{Alg:ConRes}, which demonstrates multi-agent task allocation and control strategy updates for each agent at every time step. For a more detailed algorithm on single-agent control synthesis, we defer to \cite{Enge2024} and the algorithm therein.

    \begin{algorithm}[ht]
		\caption{Task Allocation and Control Strategy Update} \label{Alg:ConRes}
		\begin{algorithmic}[1]
			\State Given: Normalized system \eqref{Sys} with $x^i(0), \ \forall i \in \mathcal{I}$
			\For{$k\in\{0,\cdots, N-1\}$}
			\State  \textbf{if} no new spec. \textbf{then} set $\mathbb{L}_{opt}=\emptyset$ \& \underline{go to line \ref{alg:linempc}}\vspace{.6em}
			\State  \textit{Task Allocation:}
			\State ($\psi$, $r_{\psi,\max}$) $\gets$ load new specification
            \State Decompose into ($\psi^j$, $r_{\psi^j,\max}$), $j \in \{1, \cdots, \nu\}$
            \State Filter pairs to obtain $\mathbb{L} \subset \mathcal{I} \times \{1, \cdots, \nu\}$
            \For{$(i,j) \in \mathbb{L}$}
            \State Solve \eqref{Eq:TMPC}, get $r^i_{\psi^j,mpc}$ and control strategy $\boldsymbol f^{(i,j)}$
            \State  \textbf{if} no solution \textbf{then} $\mathbb{L}=\mathbb{L}\setminus \{(i,j)\}$
            \EndFor
            \State Solve \eqref{Eq:OptiPair} to get definitive assignment list $\mathbb{L}_{opt}$ \vspace{.6em}
            \State  \textit{Control Strategy Update:}
			\State $\forall (i,j) \in \mathbb{L}_{opt}$, set $\boldsymbol f^{i,k}=\boldsymbol f^{(i,j)}$ $\gets$ update
            \For{$i \in \mathcal{I} \setminus \proj_{\mathcal{I}}(\mathbb{L}_{opt})$} \label{alg:linempc}
			\State Solve \eqref{Eq:TMPC} to get control strategy $\boldsymbol f^{i}_{meas}$
            \State  \textbf{if} solution \textbf{then} set $\boldsymbol f^{i,k}=\boldsymbol f^{i}_{meas}$ $\gets$ update
            \State  \textbf{else} set $\boldsymbol f^{i,k}=\boldsymbol f^{i,k-1}$ $\gets$ no update
            \EndFor
			\State Implement $\boldsymbol f^{\mathcal{I},k}$ and measure $x^i(k+1), \ \forall i \in \mathcal{I}$ \label{alg:linempc2}
			\EndFor
		\end{algorithmic}
	\end{algorithm}
 
    Let us consider the recursive feasibility of updating a control strategy for each agent at any time.
    \begin{theorem}[Recursive feasibility]
        If there exists a feasible control strategy $\boldsymbol f^{\mathcal{I},k}$, at time $k$, then there exists a feasible control strategy $\boldsymbol f^{\mathcal{I},k+1}$ at time $k+1$.
    \end{theorem}
    \begin{proof}
        It is sufficient to prove that the existence of $\boldsymbol f^{i,k}$ implies the existence of $\boldsymbol f^{i,k+1}$ for each agent $i \in \mathcal{I}$. Notice that if an agent either accepts a new sub-specification or measurement only, an updated control strategy is obtained, respectively, as $\boldsymbol f^{i,k+1}=\boldsymbol f^{(i,j)}$ or $\boldsymbol f^{i,k+1}=\boldsymbol f^{i}_{meas}$. If neither is available, the previous control strategy is again implemented, $\boldsymbol f^{i,k+1}=\boldsymbol f^{i,k}$, completing the proof.
    \end{proof}

    For each specifications $\phi$ accepted at time $k$, by subgroup $I \subseteq \mathcal{I}$, we need that $\mathbb{P}_{\boldsymbol f}(\phi, k) \geq 1-r_{\phi, \max}$. Here, it is sufficient to show that $\forall i \in I$, $\mathbb{P}_{\boldsymbol f}(\phi^i, k) \geq 1-r_{\phi^i, \max}$. This, in relation to optimization problem \eqref{Eq:TMPC}, is discussed within open-loop and closed-loop implementation in our previous work \cite{Enge2024}. In said paper, we established the validity of open-loop implementation and suggested that closed-loop implementation follows naturally from the convex unimodality of the disturbance.

    \begin{remark}
        Note that the local risk value $r_{\phi^j, mpc}$ compared to the average maximal risk $r_{\phi^j,\max}$ is conservative, $r_{\phi^j, mpc} \geq r_{\phi^j,\max}$, due to a union-bound argument used in proving \cite[Thm. 4]{Enge2024}. Also, the obtained task allocation may not always be optimal or feasible due to our heuristic choice of filtering, even if an allocation exists.
    \end{remark}

	%%%%%%%%%%%%%%%%%%%%%%%%%%%%%%%%%%%%%%%%%%%%%%%

	\section{Case Study} \label{Sec:App}

\begin{figure}[htp]
		\includegraphics[width=0.86\columnwidth]{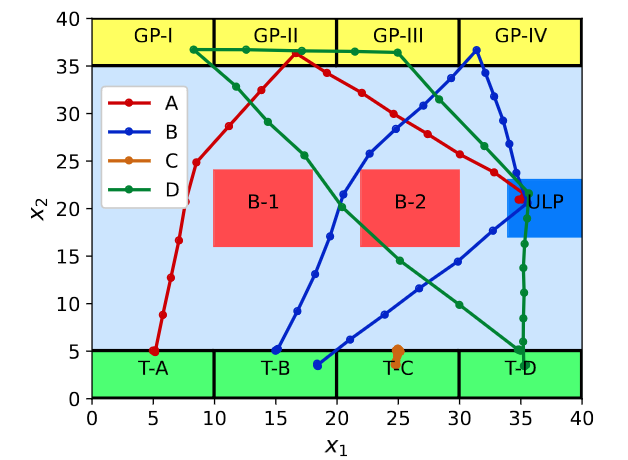}
		\centering
        \vspace{-1em}
		\caption{\small The shuttle trajectories.}
        \vspace{-1.5em}
		\label{Fig:StateMeas}
	\end{figure}

    We validate our proposed method using a multi-shuttle routing case. In a tourist attraction spot (Fig. \ref{Fig:StateMeas}), four shuttle buses of differing speeds (A-D) should pick up tourists from gathering points (GP) and deliver them to an unloading point (ULP) while avoiding two buildings (B). The shuttles initiate from their respective terminals (T) and should return when required. Each task should be assigned to a most probable shuttle to fetch the tourists. Thus, new routing may be assigned anytime, rendering a dynamic task allocation problem. 
    \if\mycmd0
	The technical details can be found in the Appendix.
	\else
    The technical details can be found in report \cite{Enge2024riskawaremulti}.
	\fi

    After decomposing, filtering, control synthesis, and auctioning, we obtain the results in Fig. \ref{Fig:StateMeas}. The logging information of the experiment indicates that: 1) at time $k=2$, buses `A', `B', and `D' are assigned with GP-II, GP-IV, and GP-I, respectively; 2) at time $k=8$, buses `A' and `D' are assigned with GP-II and GP-III, respectively; 3) at time $k=15$, buses `B', `C', and `D' have accepted the return specification, while shuttle `A' has rejected it. Bus `C' is not assigned any task throughout the experiment and, as a result, stays at its terminal. Since buses `A' and `D' accept additional tasks at $k=8$, our method allows agents to adjust their behaviours according to newly assigned tasks. As mentioned, shuttle `A' has rejected the return task at time $k=15$. This is due to the incapability to move long distances considering its small speed limit. Additionally, all buses are confined within the `BOX' region and kept outside the buildings, implying the compatibility of our method to explicitly assigned specifications. Part of the trajectory of shuttle `D' intersects with B-2 due to naive linear interpolation. In practice, nonlinear interpolation can guarantee critical safety, which may generate curved trajectories when navigating around the building.

    \begin{remark}
        In this experiment, the computation of control inputs by solving problem \eqref{Eq:TMPC} is distributed to each agent, leading to a computational complexity linear to the number of assigned specifications. In this sense, our proposed method allows for a comparable computational load of each agent to a conventional single-agent case, allowing it to be implemented in practical scenarios.
    \end{remark}

	%%%%%%%%%%%%%%%%%%%%%%%%%%%%%%%%%%%%%%%%%%%%%%%
    
	\section{Conclusion}\label{sec:conclu}

	This paper considers real-time task allocation and control synthesis for stochastic heterogeneous linear multi-agent systems with probabilistic satisfaction of STL specifications. The proposed method relies on the dynamic allocation of decomposed specifications to individual agents using risk measures via auctioning. The risk measures are computed using tube-based MPC on the agent level during the control synthesis. These two approaches provide a novel perspective for coordination control of multi-agent systems in dynamic environments. In future work, we will improve the risk quantification for less conservative control solutions.

	%%%%%%%%%%%%%%%%%%%%%%%%%%%%%%%%%%%%%%%%%%%%%%%

	\bibliographystyle{IEEEtran}
    \bibliography{Article.bib}

    \if\mycmd0

    \section*{Appendix}

    \subsection{Problem Statement: Illustrative example}

    We consider an example (see Fig. \ref{Fig:AgentSpec}) to further illustrate the problem and approach covered within this paper. %Next, we give a problem description and a problem statement. Lastly, we elaborate upon the approach as seen in Fig. \ref{Fig:Approach1}.
    
    \noindent \textbf{Example 1.}
    {\itshape We examine a multi-robotic scenario in which robots are assigned joint tasks. Upon receiving a new task, only its description and the required number of robots are known. A procedure decides which group of robots, if any, will perform the task. Given the assumption that all robots' behaviour is independent, new tasks are initially decomposed into sub-tasks at the individual level; that is, each robot performs part of the task without requiring information from other robots. Subsequently, the procedure prompts potential robot sub-task combinations to investigate an update to the individual control strategy without disrupting ongoing sub-tasks. Among all combinations, the optimal assignment is selected and all control strategies are updated accordingly.}

    \begin{figure}[htp]
        \centering
		\includegraphics[width=\columnwidth]{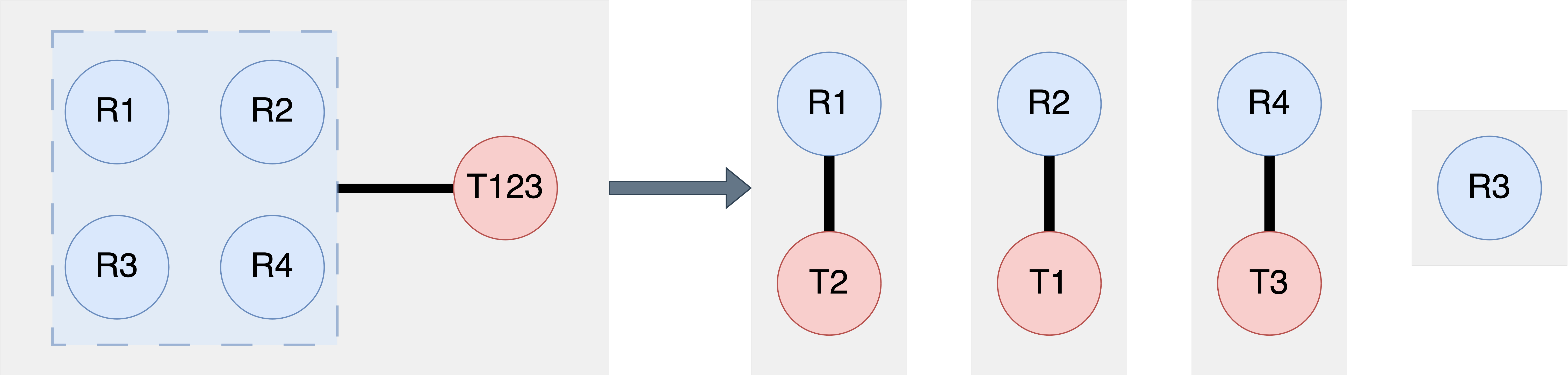}
        \vspace{-1em}
        \caption{Illustration of decomposing tasks to the individual level and combining tasks with robots. Blue are robots and red are tasks.}
		\label{Fig:AgentSpec}
	\end{figure}
   
    % \red{This is not an intuitive example as we discussed. It is a nice piece of text though.... 
    % I was thinking along the line of. 
    % Suppose that we have two delivery robots. We can describe the task to pick up mail and drop it off  as STL specifications. 
    % Over time ... 
    % We can formalize this as the following problem statement. 
    % }

    %%%%%%%%%%%%%%

    \subsection{Specification Decomposition: Illustrative example}

    We consider an example (see Fig. \ref{fig:example_dec}) to illustrate the spe-cification decomposition approach covered within this paper.

    %% example for projection onto agent's space
    \noindent \textbf{Example 2:} 
    {\itshape Consider formulas $\phi=\square_{[a,b]}\mu$ and $\psi=\square_{[a,b]}\neg\mu$, where $\mu:=(h(x)\geq \bm{0})$ is a predicate, with $h(x)=\begin{bmatrix}
       c_1 & c_2
    \end{bmatrix}x+\bm{1}_4$, $c_1=\begin{bmatrix}
        -\bm{1}_2^\intercal & \bm{1}_2^\intercal
    \end{bmatrix}^\intercal$, $c_2=\begin{bmatrix}
        1 & -1 & 1 & -1
    \end{bmatrix}^\intercal$. We wish to convert $\phi$ and $\psi$ into conjunctive forms of two conjuncts (subformulas). First, we define the polytope $H=\{x\mid h(x)\geq \bm{0}\}$, and construct orthotopes $B_{\mathrm{in}}(H)=\{x\mid b_{\mathrm{in}}(x)\geq \bm{0}\}$ and $B_{\mathrm{out}}(H)=\{x\mid b_{\mathrm{out}}(x)\geq \bm{0}\}$, with $b_{\mathrm{in}}(x)=2\begin{bmatrix}
        \bm{I}_2 & -\bm{I}_2
    \end{bmatrix}^\intercal x+\bm{1}_4$ and $b_{\mathrm{out}}(x)= \begin{bmatrix}
        \bm{I}_2 & -\bm{I}_2
    \end{bmatrix}^\intercal x+\bm{1}_4$, respectively, as in Definition \ref{def:orthotopes}. Polytope $H$ and orthotopes $B_{\mathrm{in}}(H)$, $B_{\mathrm{out}}(H)$ are depicted in Fig. \ref{fig:example_dec}. Then, using Lemma \ref{lemma:decomposition}, we define predicates $\mu_{\mathrm{in}}^i:=(x_i\in \proj_{\mathbb{X}^i}(B_{\mathrm{in}}(H))=(-0.5 \leq x_i \leq 0.5)$, $\mu_{\mathrm{out}}^i:=(x_i\in \proj_{\mathbb{X}^i}(B_{\mathrm{out}}(H))=(-1 \leq x_i \leq 1)$, with $i\in \{1,2\}$, allowing us to decompose $\mu$ and $\neg\mu$ as $\dec(\mu,2)=\mu_{\mathrm{in}}^1 \wedge  \mu_{\mathrm{in}}^2$ and $\dec(\neg\mu,2)=\neg\mu_{\mathrm{out}}^1 \wedge \neg \mu_{\mathrm{out}}^2$. Finally, $\dec(\phi,2)=\phi^1\wedge\phi^2$ and $\dec(\psi,2)=\psi^1\wedge\psi^2$, where $\phi^i= \square_{[a,b]} \mu_{\mathrm{in}}^i$ and $\psi^i= \square_{[a,b]} \neg\mu_{\mathrm{out}}^i$, with $i\in \{1,2\}$, are valid decompositions of $\phi$ and $\psi$, respectively.}
    \begin{figure}[htp]
    \centering
    \vspace{-0.8em}
    \definecolor{mycolor1}{rgb}{1.00000,0.75000,0.55000}%
    \definecolor{mycolor2}{rgb}{0.90000,0.60000,0.60000}%
    \definecolor{mycolor3}{rgb}{0.5,0.7,1.0}%
    \begin{tikzpicture}
    \begin{axis}[%
    width=1.5in,
    height=1.5in,
    scale only axis,
    xmin=-1.1,
    xmax=1.1,
    xlabel style={font=\color{white!15!black}},
    xlabel={$x_1$},
    ymin=-1.1,
    ymax=1.1,
    ylabel style=   {font=\color{white!15!black}},
    ylabel={$x_2$},
    axis background/.style={fill=white},
    axis x line*=bottom,
    axis y line*=left,
    xmajorgrids,
    ymajorgrids,
    legend style={at={(0.7,0.8)}, anchor=south west, legend cell align=left, align=left, draw=white!15!black, fill  opacity=0.7, font=\footnotesize}
    ]
    \addplot[area legend, line width=1.0pt, draw=black,     fill=mycolor3, fill opacity=0.5]
    table[row sep=crcr] {%
    x	y\\   
    1	1\\
    -1	1\\
    -1	-1\\
    1	-1\\
    }--cycle;
    \addlegendentry{$B_{\mathrm{out}}$}
    \addplot[area legend, line width=1.0pt, draw=black,     fill=mycolor1]
    table[row sep=crcr] {%
    x	y\\
    0	-1\\
    1	0\\
    0	1\\
    -1	0\\
    }--cycle;
    \addlegendentry{$H$}
    \addplot[area legend, line width=1.0pt, draw=black,         fill=mycolor2]
    table[row sep=crcr] {%  
    x	y\\
    0.5	0.5\\
    -0.5	0.5\\
    -0.5	-0.5\\
    0.5	-0.5\\
    }--cycle;
    \addlegendentry{$B_{\mathrm{in}}$}
    \end{axis}
    \begin{axis}[%
    width=2in,
    height=2in,
    at={(0in,0in)},
    scale only axis,
    xmin=0,
    xmax=1,
    ymin=0,
    ymax=1,
    axis line style={draw=none},
    ticks=none,
    axis x line*=bottom,
    axis y line*=left
    ]
    \end{axis}
    \end{tikzpicture}%
        \vspace{-1em}
        \caption{Polytopes $H$, $B_{\mathrm{in}}(H)$, $B_{\mathrm{out}}(H)$, for Example 2-.}
        \label{fig:example_dec}
        \vspace{-1em}
    \end{figure}

    %%%%%%%%%%%%%%

    \subsection{Case Study: Technical Details}

    We select a control horizon $N = 25$ and use a polytopic predicate $\mu_{\mathrm{R}} := (x_k \in \mathrm{R})$ to denote the position $x_k$ of a shuttle residing within a rectangular region $\mathrm{R}$. 
    %In addition to global and local specifications investigated in this paper, the scenario also allows for specifications explicitly assigned to individual agents, such as routing a specific bus to a designated region. 
    The overall task specifications are listed as follows.  
    \begin{itemize}
    \item Initially, each shuttle $X$ should satisfy a safety specification $\varphi_{(0,X)} \!:=\! \square_{[0, N]} (\lnot \mu_{\mathrm{B}\textbf{-}1} \wedge \lnot \mu_{\mathrm{B}\textbf{-}2})$.
    \item A global task is assigned at $k=2$: \textit{the shuttles should deliver tourists from GP-I,\,II,\,IV to ULP within 15 time steps}. We decompose it into three local specifications: \\
    $\varphi_{(2,\mathrm{I})}\!:=\!\lozenge_{[2, 9]} \mu_{\mathrm{GP}\textbf{-}\mathrm{I}} \!\wedge\! \square_{[2, 9]} ( \mu_{\mathrm{GP}\textbf{-}\mathrm{I}} \!\rightarrow\! \lozenge_{[0, 7]} \mu_{\mathrm{ULP}})$, \\
    $\varphi_{(2,\mathrm{II})}\!:=\!\lozenge_{[3, 10]} \mu_{\mathrm{GP}\textbf{-}\mathrm{II}} \!\wedge\! \square_{[3, 10]} ( \mu_{\mathrm{GP}\textbf{-}\mathrm{II}} \!\rightarrow\! \lozenge_{[0, 7]} \mu_{\mathrm{ULP}})$, \\
    $\varphi_{(2,\mathrm{IV})}\!:=\!\lozenge_{[4, 11]} \mu_{\mathrm{GP}\textbf{-}\mathrm{IV}} \!\wedge\! \square_{[4, 11]} ( \mu_{\mathrm{GP}\textbf{-}\mathrm{IV}} \!\rightarrow\! \lozenge_{[0, 7]} \mu_{\mathrm{ULP}})$.
    \item A global task is assigned at $k=8$: \textit{the shuttles should deliver tourists from GP-II,\,III to ULP within 15 time steps}. We decompose it into two local specifications: \\
    $\varphi_{(8,\mathrm{III})}\!:=\!\lozenge_{[8, 15]} \mu_{\mathrm{GP}\textbf{-}\mathrm{III}} \!\wedge\! \square_{[8, 15]} ( \mu_{\mathrm{GP}\textbf{-}\mathrm{III}} \!\rightarrow\! \lozenge_{[0, 7]} \mu_{\mathrm{ULP}})$, \\
    $\varphi_{(8,\mathrm{II})}\!:=\!\lozenge_{[9, 16]} \mu_{\mathrm{GP}\textbf{-}\mathrm{II}} \!\wedge\! \square_{[9, 16]} ( \mu_{\mathrm{GP}\textbf{-}\mathrm{II}} \!\rightarrow\! \lozenge_{[0, 7]} \mu_{\mathrm{ULP}})$.
    \item At time $k=15$, each shuttle $X$ is assigned with a `back to terminal' specification $\varphi_{(15,X)}\!:=\!\lozenge_{[15, 23]} \square_{[0, 2]} \mu_{\mathrm{T}\textbf{-}\mathrm{X}}$.
    \end{itemize}
    
    All four buses have the same dynamic models described by system~\eqref{Sys} with $A\!=\!B\!=\!I_2$ and $w^i(k) \sim \mathcal{N}(0,\sigma I_2)$ for all $i \in \mathcal{I}$, with $\sigma=0.001$. The states and the control inputs are their positions and velocities, respectively. To incorporate the heterogeneity of practical systems, we assume that the buses have different speed limits, i.e., $4\,$m/s, $5\,$m/s, $6\,$m/s, and $7\,$m/s for buses `A' to `D', respectively. We consider a stabilizing feedback gain $K\approx -0.618 I_2$ for Eq.~\eqref{ErrorPart}, a maximal risk $r_{*, \max}=0.5$ for any global specification or explicit specification ($k=0$, $k=15$), individual maximal risk based on $r_{\phi^j,\max}=\frac{r_{\phi,\max}}{\nu}$ for any local specification ($k=2$, $k=8$) and a cost $J(\boldsymbol{z}, \boldsymbol{v}, \boldsymbol{r}) = u^TRu +\sum_{i=1}^N r(i)$ for Eq. \eqref{Eq:TMPC_cost} with $R=0.001I_2$. %A coordinate transformation ensures all PRS are spherical and the dynamics are normalized.

    Similar to \cite{Enge2024}, instead of the non-linear constraints in Eq. \eqref{Eq:TMPC_ConRiskPRT}, we impose the following equality and inequality constraints, implying that the original constraints hold. These constraints are computationally preferable, as they are quadratic convex equality and linear inequality constraints. We replace the non-linear equality \eqref{Eq:TMPC_ConRiskPRT} with a quadratic equality $r(k) = a \rho(k)^2 +b$ and a linear inequality $0.01 \leq r(k) \leq 1$, with $a=-0.005$ and $b=1.01$. For $r(k) \in [\,0.01, 1\,]$, one can verify that $a \rho(k)^2 +b \geq \frac{n}{\rho(k)^2}$ always holds. Thus, the transformed optimization problem provides a sound solution to obtaining local risk values and controllers.

	\else
	\fi

	\end{document}